\documentclass[5p]{elsarticle}
\usepackage[T1]{fontenc}
\usepackage{amsmath}
\usepackage{graphicx}
\usepackage{lmodern}
\usepackage[latin9]{inputenc}
\usepackage{hyperref}
\usepackage{bm}

\newcommand{\eqnref}[1]{Eq.~\eqref{#1}}

\newcommand{\figref}[1]{Fig.~\ref{fig:#1}}

\newcommand{\bo}[1]{\mathbf{#1}}

\newcommand{\secref}[1]{Section \ref{#1}}
\begin{document}

\title{Scaling of Coercivity in a 3d Random Anisotropy Model}

\author[cor]{T.~C.~Proctor}
\ead{proctortc@gmail.com}
\author{E.~M.~Chudnovsky}
\ead{EUGENE.CHUDNOVSKY@lehman.cuny.edu}
\author{D.~A.~Garanin}
\address{Physics Department, Lehman College, City University of New York \\
 250 Bedford Park Boulevard West, Bronx, New York 10468-1589, USA}
\cortext[cor]{Corresponding author, phone: +1-203-339-2230}

\date{\today}

\begin{abstract}
The random-anisotropy Heisenberg model is numerically studied  on lattices containing over ten million spins. 
The study is focused on hysteresis and metastability due to topological defects, and is relevant to magnetic properties of amorphous and sintered magnets. 
We are interested in the limit when ferromagnetic correlations extend beyond the size of the grain inside which the magnetic anisotropy axes are correlated. 
In that limit the coercive field computed numerically roughly scales as the fourth power of the random anisotropy strength and as the sixth power of the grain size. 
Theoretical arguments are presented that provide an explanation of numerical results. 
Our findings should be helpful for designing amorphous and nanosintered materials with desired magnetic properties. 
\end{abstract}


\maketitle

\section{Introduction}
The random anisotropy model was introduced by Harris, Plischke, and Zuckermann\cite{harplizuc73prl} to describe magnetic properties of amorphous ferromagnets. 
The problem is subtle when local magnetic anisotropy is weak compared to the exchange interaction, which is usually the case due to the relativistic nature of the anisotropy. 
In this case the exchange interaction creates extended ferromagnetic ordering. 

In a crystalline ferromagnet the ordered regions would correspond to ferromagnetic domains separated by thin domain walls, with the magnetization inside the domains aligned with the directions of the global anisotropy axes. 
If one neglects the magnetic dipole interaction, the ground state corresponds to the infinite size of the domain, that is, to the global ferromagnetic ordering. 
It is the magnetic dipole interaction that breaks ferromagnetic crystal into domains, with the ground state corresponding to  zero total magnetic moment. 
In practice, however, pinning of domain walls by disorder results in the magnetic hysteresis that permits permanent magnets. 

In a random-anisotropy ferromagnet the global directions of the anisotropy are absent. 
As one moves along a certain direction in a solid, the magnetization vector, created by the exchange, feels weak random kicks from the magnetic anisotropy. 
This effect resembles a random walk. 
At large distances, it results in a significant deviation of the magnetization from its original direction, thus creating magnetic domains of a different nature. 
This problem was first analyzed by Imry and Ma\cite{Imry-Ma-PRL1975,CT-book} in the general context of a vector field in $d$ dimensions interacting with a weak random field. 
They argued that a random field, no matter how weak it may be, destroys the long range order in less than four dimensions. 

This concept was further developed by a number of authors\cite{PelcovitsPRL,PelcovitsPRB,Aharony,Feldman,Fedorenko} who used the functional renormalization group and replica-symmetry-breaking methods in application to random-anisotropy systems. 
A phenomenological approach had been also suggested\cite{CSS,Chudnov1983,CS-JMMM,CSPRB} along the lines of the Imry-Ma argument which explained some of the vast amount of experimental data on amorphous ferromagnets.\cite{amorphous-book} 
The same concept was applied to other systems, such as superfluid 3He-A in aerogel.\cite{Volovik} 
There has also been analytical work accompanied by Monte Carlo studies on small lattices\cite{Fisch,Itakura,Imagawa,Dudka} which assumed thermal equilibrium. 

Some of the conclusions of these studies came into question after experiments, as well as the early numerical work,\cite{Serota,Barbara,DC} indicated that random-anisotropy systems exhibited metastability and glassy behavior\cite{Billoni2005,Billoni2008} which is not captured by the Imry-Ma argument or the replica symmetry breaking method. 
This was confirmed by recent numerical work on the three dimensional random field $XY$ model on lattices of up to one billion spins.\cite{XYPRB,XYEPL} For any practical purpose, this makes random field and random anisotropy systems not very different from a conventional ferromagnet, having high metastability and magnetic hysteresis that only disappears at sufficiently high temperature or exponentially long times. 

Although it was long suspected that random-field systems formed some kind of a ``vortex glass'', the mechanisms of metastability was poorly understood. 
Most recently, however, a topological argument has been proposed\cite{topology-PRL} and confirmed by large-scale numerical computations showing that reversible Imry-Ma type behavior of the random-field spin system only emerges when the number of spin components $n$ exceeds $d + 1$, with $d$ being the dimensionality of space. 
At $n < d +1$, the formation of the Imry-Ma state requires topological defects, such as hedgehogs in the $n = d = 3$ Heisenberg model, which leads to metastability. 
A similar argument applies to the random-anisotropy model, although there are peculiarities which we will discuss in \secref{discussion}. 

In this paper we study the random anisotropy Heisenberg model on lattices in excess of ten million spins. 
The emphasis is on measurable quantities, such as magnetic hysteresis, the coercive field, and their dependence on the anisotropy strength and the size of the volume inside which the anisotropy axes are correlated. 
The latter is relevant to the magnets sintered from randomly oriented nanoscopic ferromagnetic grains. 
We find very strong dependence of the magnetic properties on  parameters, which we believe is important for synthesizing materials with desired magnetic properties.

The paper is structured as follows: 
The model is formulated in \secref{model}, where some analytical results are also obtained. 
The numerical method and results are presented in  \secref{results}.  We begin by analyzing the case of site disorder.  \secref{correlations} compares short-range correlations computed numerically with analytical results, and provides spin-spin correlation functions for different initial conditions.  
\secref{hysteresis} presents computed hysteresis curves and obtains their scaling with the strength of the random anisotropy. 
The role of hedgehogs in the magnetic state is discussed in  \secref{hedgehogs}.  
\secref{sintered} presents numerical results and their interpretation in the random anisotropy system that has short range correlations in the distribution of the anisotropy axes.  
\secref{discussion} contains some final remarks and suggestions for experiment. 

\section{Model and Analytical Results}\label{model}
The three-dimensional Heisenberg model with random anisotropy is described by the 
Hamiltonian
\begin{equation}
\mathcal{H}=-\frac{J}{2}\sum_{i,j}\mathbf{s}_i\cdot {\bf s}_j
-D_R \sum_i ({\bf n}_i \cdot {\bf s}_i)^2
-{\bf H} \cdot \sum_i {\bf s}_i, \label{Hamiltonian}
\end{equation}
where the first sum is over nearest neighbors,
${\bf s}_i$ is a three component spin of constant length $s$,
${\bf H}$ is the external field,
$D_R$ is the strength of the random anisotropy, and
${\bf n}_i$ is a three-component unit vector having random direction at each lattice site. We assume ferromagnetic exchange, $J > 0$. 
The factor of $1/2$ in front of the first term is needed to count the exchange interaction $Js^2$ between each pair of spins once. 
In our numerical work we consider a cubic lattice. 
For the real atomic lattice of cubic symmetry the single-ion anisotropy of the form $-({\bf n} \cdot {\bf s})^2$ would be absent, the first non-vanishing anisotropy terms would be of the form $s_x^2s_y^2+s_x^2s_z^2+s_y^2s_z^2$. 
However, in our case the choice of a cubic lattice is merely a computational tool that should not affect our conclusions. 

In a cubic lattice the effective exchange field acting on each spin is $6Js$ due to six nearest neighbors. 
In our model it competes with the anisotropy field of strength $2sD_R $. 
The case of a large random anisotropy, $2sD_R  \gg 6Js$, that is, $D_R \gg 3J$, is obvious, corresponding to a system of weakly interacting randomly oriented single-domain particles. 
At $T = 0$ each spin aligns with the local ${\bf n}$. 
At $T = 0$, due to the two equivalent directions along the easy axis, the system possesses magnetic hysteresis with a coercive field, $H_C$, of the order of the local anisotropy field $H_C \propto 2sD_R$. 

Here we are interested in a more subtle case of  weak random anisotropy, $D_R \ll 3J$. 
Such anisotropy cannot destroy the local ferromagnetic order created by the strong exchange interaction, it can only slightly disturb the direction of the magnetization when one goes from one lattice site to the other. 
If one moves along a line in a solid such random disturbances would resemble a random walk. 
Consequently the deviation of the direction of the magnetization would grow with the distance. 
In a three-dimensional lattice of spacing $a$ the average statistical fluctuation of the random anisotropy field per spin in a volume of size $R$ would scale as $D_{\rm eff} = 2sD_R (a/R)^{3/2}$, while the ordering effect of the exchange field would scale as $6Js(a/R)^2$. 
They become comparable at $R/a = R_f/a \sim (3J/D_R)^2$. 

This famous argument\cite{Imry-Ma-PRL1975} provides an estimate of the size of the Imry-Ma domain, i.e.~the distance $R_f$ on which the magnetization rotates by a significant angle. 
It leaves open the question whether the ground state of the random anisotropy system possesses a non-zero magnetization $M$. 
Even if it does, as is the case in the domain state of a conventional macroscopic ferromagnet, the state with $M = 0$ may have no practical significance because the presence of topological defects and their pinning by disorder will always result in metastability and magnetic hysteresis. 
The coercive field in the weak random anisotropy case must be proportional to $D_{\rm eff}$ on the scale $R_f$, which gives $H_C \propto D_R^4/J^3$. 
The proportionality of $H_C$ to the fourth power of $D_R$ gives a very soft magnet in the limit of small $D_R$. This can be extended to the limit of a Heisenberg ferromagnet with no anisotropy at all, which has infinite susceptibility. 

The qualitative arguments presented above can be refined using a continuous field theory version of the Hamiltonian given in \eqnref{Hamiltonian}, 
\begin{equation}
{\cal H}=\int d^{3}r\left[\frac{\alpha}{2}(\partial_{\mu}{\bf S})\cdot(\partial_{\mu}{\bf S})-\frac{\beta_R}{2} ({\bf n} \cdot {\bf S})^2 -{\bf H}\cdot{\bf S}\right],\label{H-continuous}
\end{equation}
where $\alpha = Ja^5$, $\beta_R = 2D_R a^3$, ${\bf S}({\bf r})$ is a three-component spin field of length $S_0=s/a^3$, and ${\bf n}({\bf r})$ is a three-component random field of unit length.  
Adding  a Lagrange multiplier term,  $- \int d^3 r \,\lambda({\bf r}) {\bf S}^2$, to \eqnref{H-continuous} which accounts for the  ${\bf S}^2({\bf r})$ being constant, one obtains the following equation for the extremal ${\bf S}({\bf r})$ configurations:
\begin{equation}
\alpha \nabla^2 {\bf S} + \beta{\bf n}({\bf n}
\cdot {\bf S}) + 2\lambda {\bf S} = 0
\end{equation}
Multiplying by ${\bf S}_0$, one obtains an equation for $\lambda$. 
At $R \ll R_f$ it gives
\begin{equation}
\alpha \nabla^2 {\bf S} = - \beta{\bf n}({\bf S}\cdot {\bf n}) + \frac{\beta}{S_0^2}{\bf S}({\bf S}\cdot {\bf n})^2
\end{equation}
\begin{multline}
{\bf S}({\bf r}) = -\frac{\beta}{\alpha} \int d^3r'G({\bf r} - {\bf
r}') \times 
 \bigg\{{\bf n}({\bf r}')[{\bf S}({\bf r}')\cdot {\bf n}({\bf r}')]\\
\left.-\frac{1}{S_0^2}{\bf S}({\bf r}')[{\bf S}({\bf r}')\cdot {\bf n}({\bf r}')]^2\right\}
\end{multline}
where $G({\bf r}) = -1/(4\pi |{\bf r}|)$ is the Green function of the Laplace equation. 
Then
\begin{align}\label{S-n}
& \frac{1}{2S_0^2}\langle[{\bf S}({\bf r}_1) - {\bf S}({\bf r}_2)]^2\rangle = 1 - \frac{1}{S_0^2}\langle{\bf S}({\bf r}_1)\cdot
{\bf S}({\bf r}_2)\rangle =\nonumber \\
& =\frac{\beta^2}{2\alpha^2} \int d^3r' \int d^3r''[G({\bf r}_1 -
{\bf r}')-G({\bf r}_2 - {\bf r}')] \nonumber \\
& \times [G({\bf r}_1 -
{\bf r}'')-G({\bf r}_2 - {\bf r}'')]\langle{\bf u}({\bf r}')\cdot
{\bf u}({\bf r}'')\rangle 
\end{align}
where ${\bf u} =  {\bf n}({\bm \sigma} \cdot {\bf n}) - {\bm \sigma}({\bm \sigma}\cdot {\bf n})^2$. 

At $R_f \gg a$ the direction of ${\bf S}$ is roughly uncorrelated with the direction of ${\bf n}$ at the same lattice site. 
This gives
\begin{align}\label{nS-corr}
&\langle{\bf u}({\bf r}')\cdot {\bf u}({\bf r}'')\rangle = \langle n'_{\alpha} n'_{\beta} n''_{\gamma} n''_{\delta}\rangle \times \nonumber \\
&  \langle \sigma'_{\beta} \sigma''_{\delta}(\sigma'_{\alpha}\sigma'_{\mu} -\delta_{\alpha \mu})(\sigma''_{\gamma}\sigma''_{\mu}-\delta_{\gamma\mu})\rangle,
\end{align}
where ${\bf n}' = {\bf n}({\bf r}'), {\bf n}'' = {\bf n}({\bf r}'')$ and the same for ${\bm \sigma}$.
The general form of the anisotropy correlator is 
\begin{align}\label{n-corr}
&\langle n'_{\alpha} n'_{\beta} n''_{\gamma} n''_{\delta}\rangle =  \frac{1}{5}[A({\bf r}',{\bf r}'')\delta_{\alpha\beta}\delta_{\gamma\delta}\\ \nonumber
&+ B({\bf r}',{\bf r}'')(\delta_{\alpha\gamma}\delta_{\beta\delta} + \delta_{\alpha\delta}\delta_{\beta\gamma})]
\end{align}
The condition ${\bf n}^2 = 1$ gives $A = B = 1$ at ${\bf r}' = {\bf r}''$, and $A = 5/3$, $B = 0$ at $|{\bf r}' - {\bf r}''| \rightarrow \infty$. 
It is easy to see that the $A$-term in \eqnref{n-corr} does not contribute to \eqnref{nS-corr}. 
Replacing $B$ with  $a^3\delta({\bf r}' - {\bf r}'')$, one obtains
\begin{align}
& \frac{1}{2S_0^2}\langle[{\bf S}({\bf r}_1) - {\bf S}({\bf r}_2)]^2\rangle = 1 - \frac{1}{S_0^2}\langle{\bf S}({\bf r}_1)\cdot
{\bf S}({\bf r}_2)\rangle =\nonumber \\
& =\frac{\beta^2a^3}{15\alpha^2} \int d^3r[G({\bf r}_1 -
{\bf r})-G({\bf r}_2 - {\bf r})]^2 \nonumber \\
& = \frac{\beta^2a^3}{60\pi\alpha^2} |{\bf r}_1 - {\bf r}_2| = \frac{|{\bf r}_1 - {\bf r}_2|}{R_f}
\end{align}
at ${|{\bf r}_1 - {\bf r}_2|} \ll {R_f}$, where 
\begin{equation}\label{Rf}
\frac{R_f}{a} =\frac{60\pi \alpha^2}{\beta^2a^4} = 15 \pi \left(\frac{J}{D_R}\right)^2
\end{equation}
in accordance with the Imry-Ma argument. 

\section{Numerical Results}\label{results}
Our numerical method consists of two processes randomly chosen for each lattice site.
The first process, which  we call ``relaxation'', rotates the spin towards the direction of the effective field, defined by
\begin{equation}
{\bf H}_{i,\mathrm{eff}}= -\frac{\delta {\cal{H}}}{\delta {\bf s}_i} = J\sum_j {\bf s}_j + 2 D_R({\bf n}_i \cdot {\bf s}_i ){\bf n}_i + {\bf H} . \label{efieldef}
\end{equation}
The rotated spin is then ${\bf s_{i,\mathrm{new}}}=s{{\bf H}_{i,\mathrm{eff}}}/{\left|{\bf H}_{i,\mathrm{eff}}\right|}$.

In the second process, which we call ``overrelaxation'', the spin is rotated by $180^o$ about the direction of the effective field,  \eqnref{efieldef}.
The new spin is given by
\begin{equation}
 {\bf s}_{i,\mathrm{new}}=
 \frac{2\left({\bf s}_{i,\mathrm{old}}\cdot {\bf H}_{i,\mathrm{eff}}\right){\bf H}_{i,\mathrm{eff}}}{ H_{i,\mathrm{eff}}^2}-{\bf s}_{i,\mathrm{old}} 
\end{equation}
Substituting this into the original Hamiltonian, one finds that at $R_f \gg a$, i.e.~when the nearest-neighbor spins are approximately aligned,  overrelaxation reduces the energy. 
At each site, we randomly choose between the two processes, and continue do so throughout the lattice, repeating until we reach convergence. 
This overrelaxation method has been found to produce much faster convergence than the ordinary relaxation.  
The combination of relaxation and overrelaxation converges to a representative local energy minimum that is typical of a glassy system. 
All our computations are done at zero temperature and therefore are relevant to the hysteretic behavior of the random anisotropy system at temperatures well below the Curie temperature of the local ferromagnetic ordering.  

\subsection{Correlation Functions}\label{correlations}

\begin{figure}[h!]
\centering
\includegraphics[width=8.5cm]{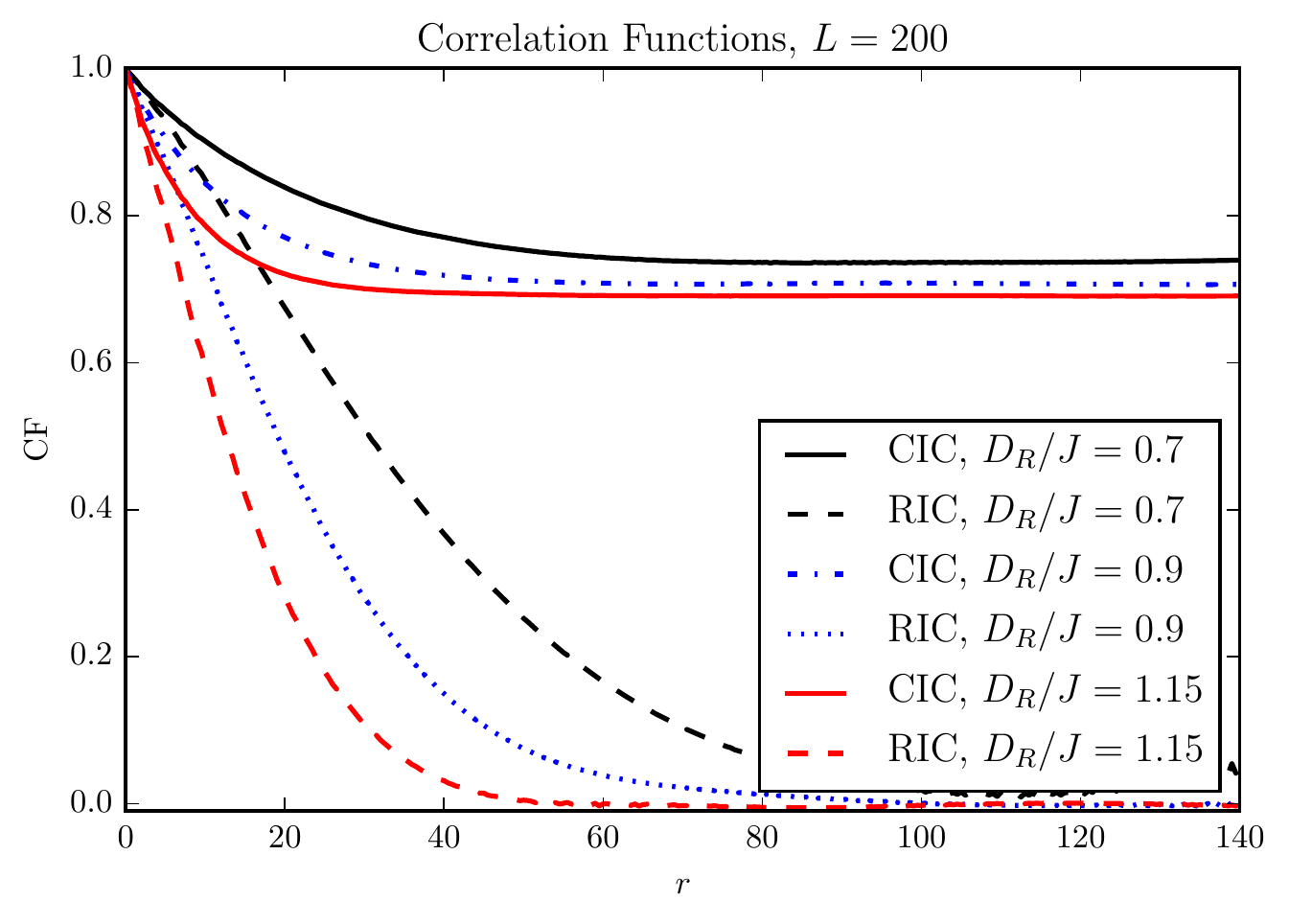}
\caption{\label{fig:CFs} Correlation functions from random initial conditions (RIC) and collinear initial conditions (CIC). Full color online. 
%\todo[inline]{round $D_R$ numbers to save space}
%\todi{Re-write title, only have one model in this paper}
}
\end{figure}
We have computed spin-spin correlation functions, defined by
$CF\left(R\right)\equiv \left<\bo{s}(\bo{r})\cdot \bo{s} \left(\bo{r} - \bo{R}\right)\right>$.
Two initial conditions have been used. 
Collinear initial conditions (CIC) physically correspond to the state obtained by placing the sample in a strong magnetic field which is then turned off. 
Random initial conditions (RIC) physically correspond to fast cooling followed by relaxation in zero magnetic field. 
Correlation functions are shown in \figref{CFs}. 
As would be expected, the curves differ significantly depending on initial conditions. 
Under collinear initial conditions, the CF levels off to a finite value, in agreement with the significant magnetization that remains. 
However, correlations go to zero for random initial conditions, consistent with zero magnetization. 

\begin{figure}[h!]
\centering
\includegraphics[width=8.5cm]{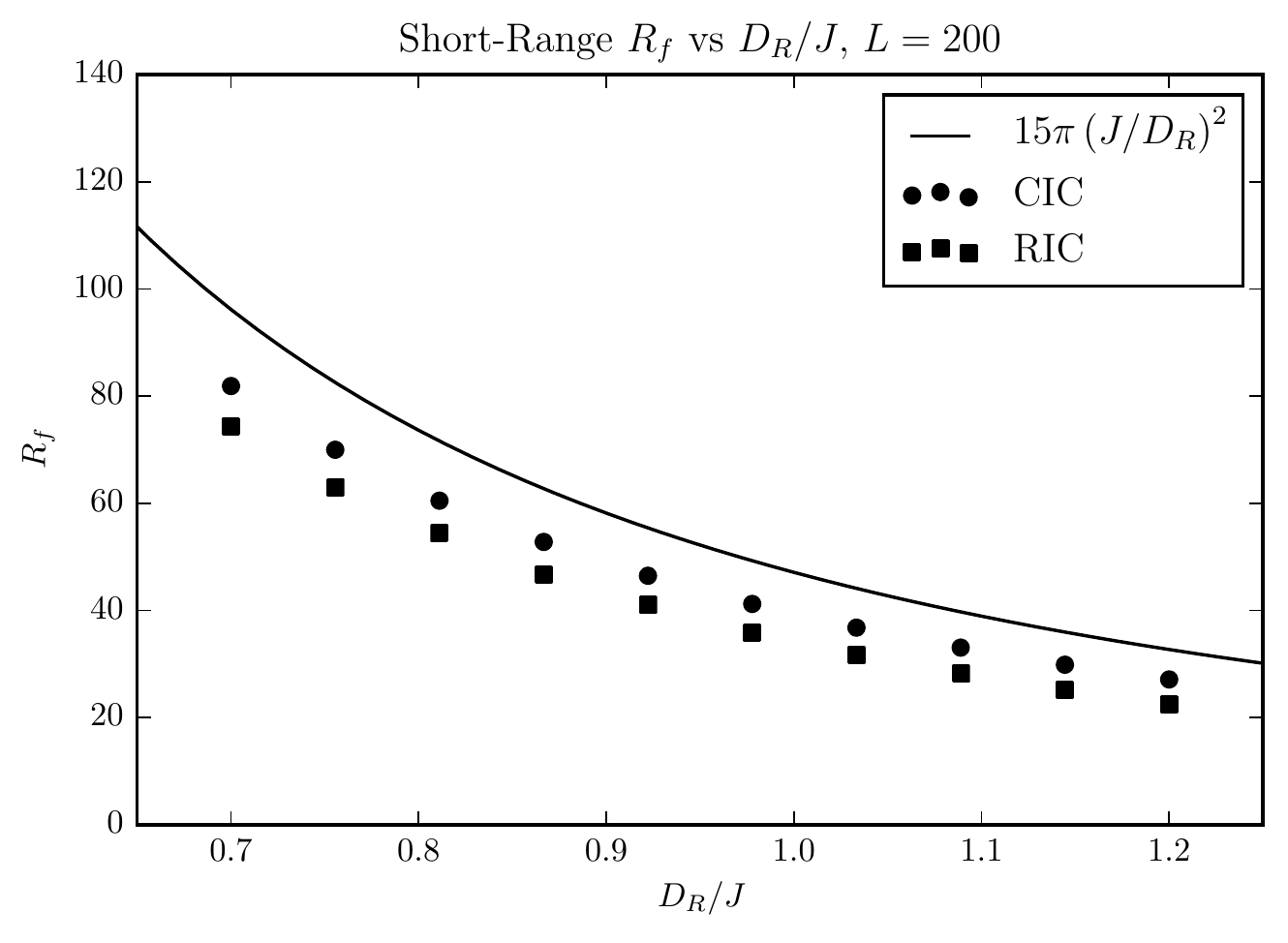}
\caption{\label{fig:ShortRangeRf}Short-range Correlation lengths for CIC and RIC }
\end{figure}
It is interesting to compare the linear decrease of the CF at very small $R$ with the prediction of the analytical theory, $CF = s^2(1-R/R_f)$. 
The dependence of $R_f$ on $D_R$ extracted from the linear dependence of the CF on $R$ at $R \ll R_f$ is shown in \figref{ShortRangeRf}. 
It is consistent with \eqnref{Rf}, although the agreement is not exact. 
This is not surprising since the analytical theory did not account for topological defects, which we discuss in \secref{discussion}. 
At greater $R$ the correlation function for the state obtained from the RIC roughly follows $\exp(-R/R'_f)$ with $R'_f$ given by ${R'_f}/{a} \approx 22 \left({J/}{D_R}\right)^2$. 
While $R'_f$ is slightly shorter than $R_f$, it also follows the $1/D_R^2$ dependence, in agreement with the Imry-Ma argument.  
\begin{figure}[h!]
\centering
\includegraphics[width=8.5cm]{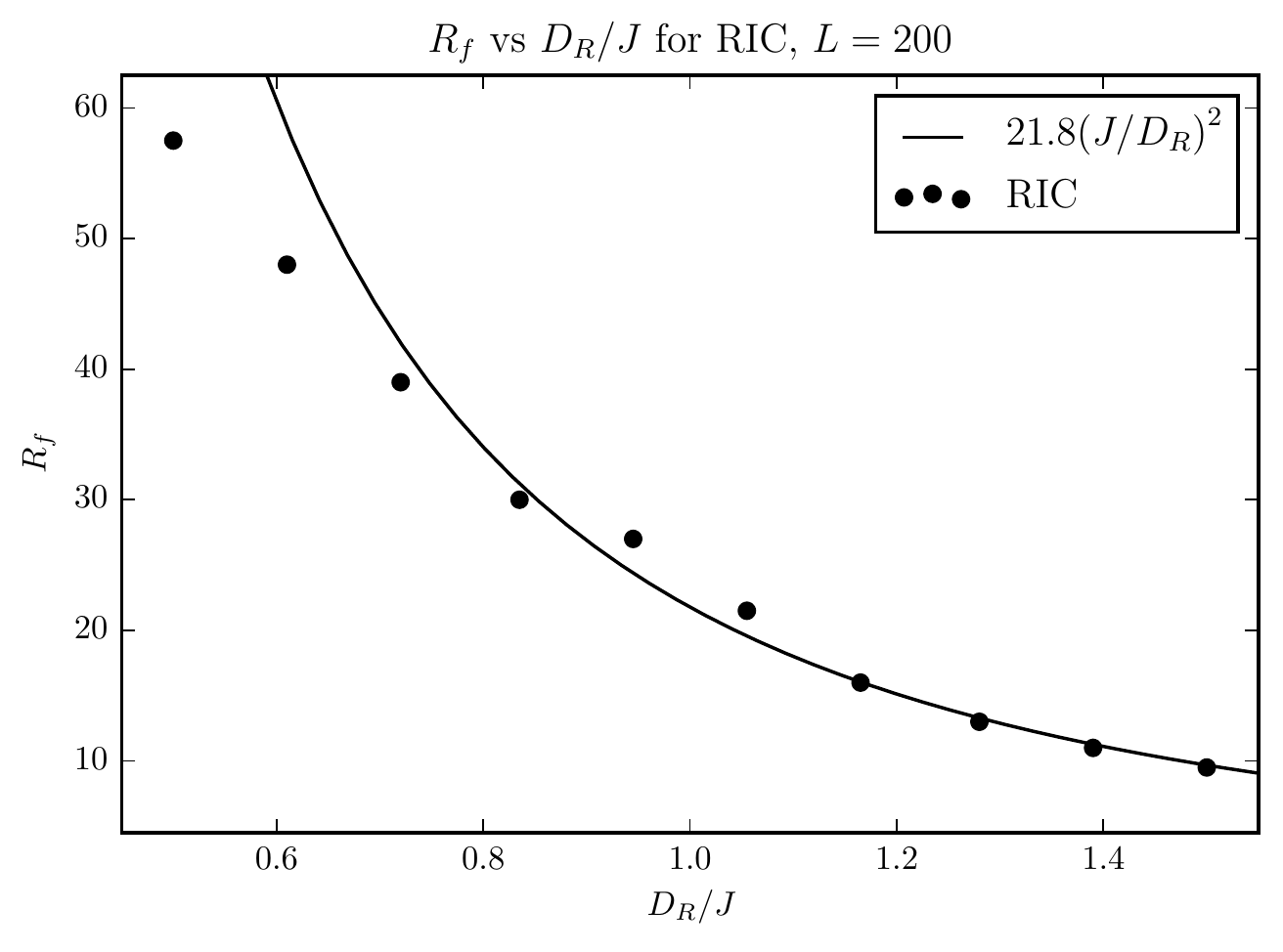}
\caption{\label{fig:RhvsDrRIC}Correlation lengths for RIC
%\todi{Re-write title, only have one model in this paper}
}
\end{figure}

\subsection{Hysteresis}\label{hysteresis}
\begin{figure}[h!]
\centering
\includegraphics[width=8.5cm]{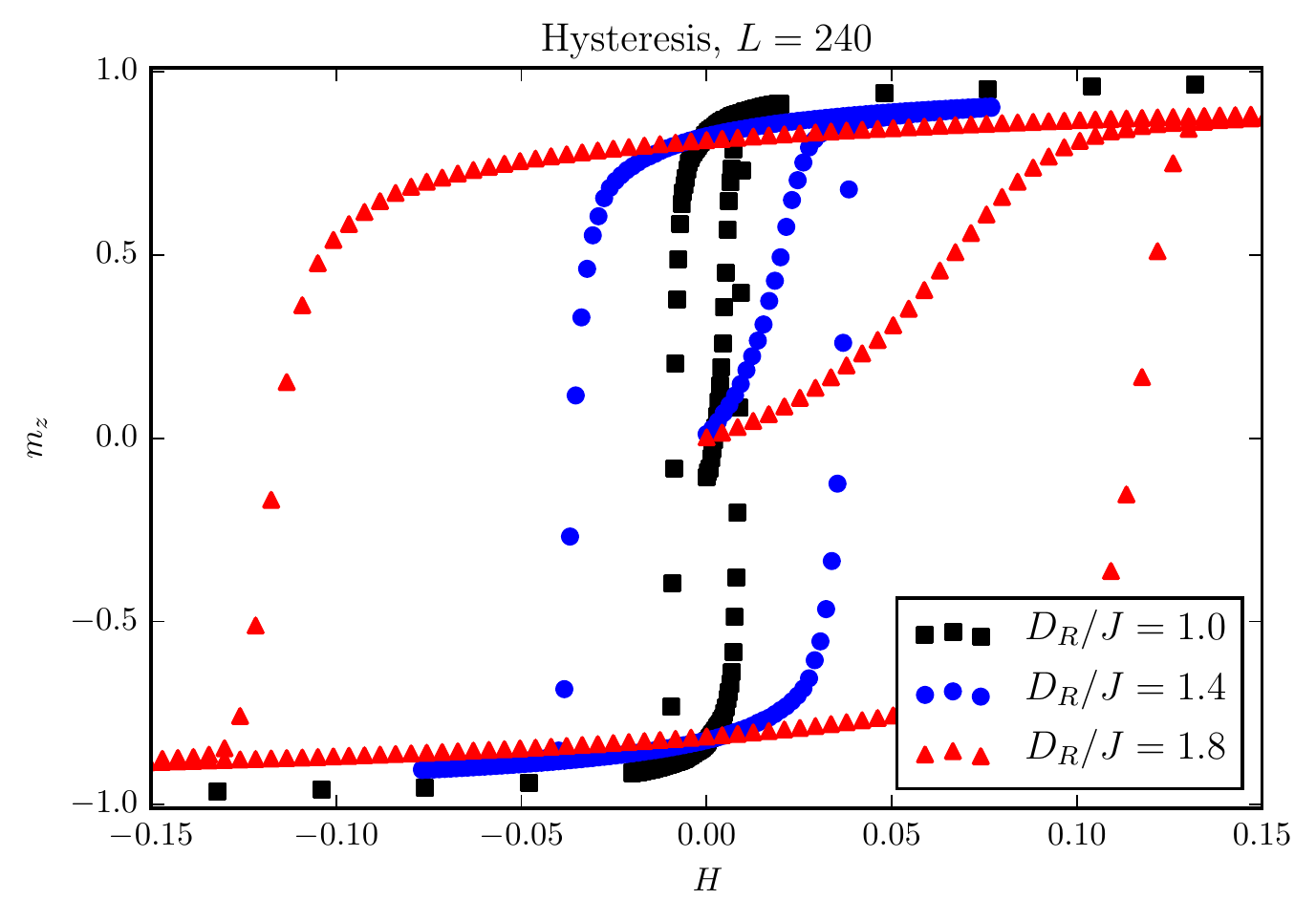}
\caption{\label{fig:HysteresisUnscaled}Hysteresis curves.  Full color online.}
\end{figure}
\begin{figure}[h!]
\centering
\includegraphics[width=8.5cm]{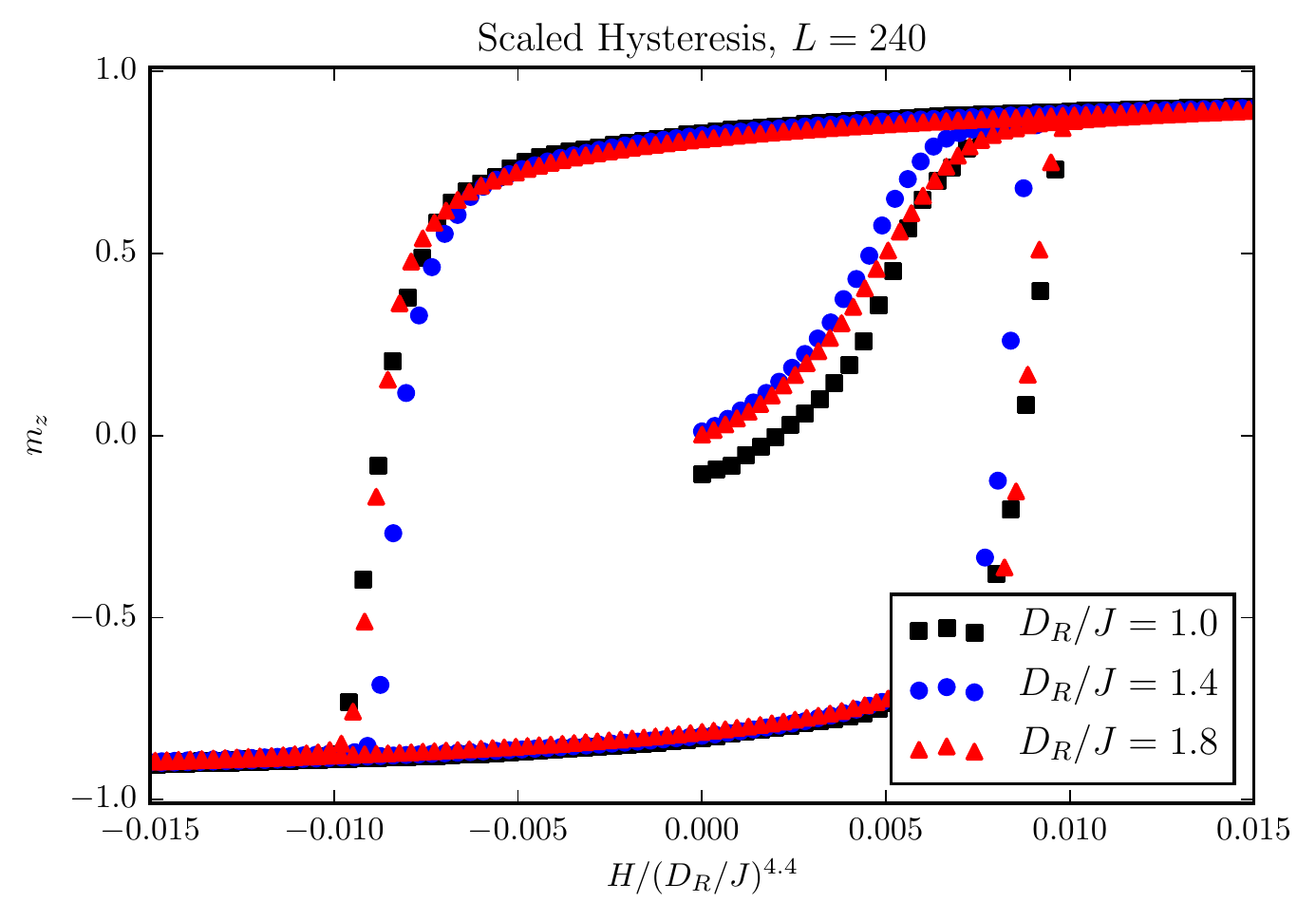}
\caption{\label{fig:HysteresisScaled}Hysteresis curves scaled.  Full color online}
\end{figure}
We have numerically computed hysteresis curves using the above method. 
The results for different $D_R$ are shown in \figref{HysteresisUnscaled}. 
They can be reasonably well scaled by dividing $H$ by a certain power of $D_R$ as is shown in \figref{HysteresisScaled}. 
This scaling allows one to approximate the coercive field, $H_C$, i.e.~the field required to bring magnetization to zero from saturation, by $H_C\approx {D_R^{4.4}}/{118}$. 
The area of the hysteresis loop scales similarly.  
This is roughly consistent with the expectation that $H_C$ scales as the fourth power of $D_R$, although the agreement is not precise.

\subsection{Hedgehogs}\label{hedgehogs}
$3d$ Heisenberg model has topological defects -- ``hedgehogs''-- which correspond to the magnetization vector field going into a point or sticking out of a point. 
Hedgehogs possess $\pm1$ topological charge and thus appear in pairs. 
In the absence of random anisotropy, hedgehogs and anti-hedgehogs would be attracted to each other and would annihilate. 
However, random anisotropy can stabilize hedgehogs even at $T = 0$. Random initial conditions automatically introduce hedgehogs. 
Relaxation from RIC annihilates some of the hedgehog pairs but leaves the system with a finite residual number of hedgehogs which depends on the strength of the random anisotropy. 
This must be one of the reasons why predictions of the continuous model deviate from numerical results. 

\begin{figure}[h!]
\centering
\includegraphics[width=8.5cm]{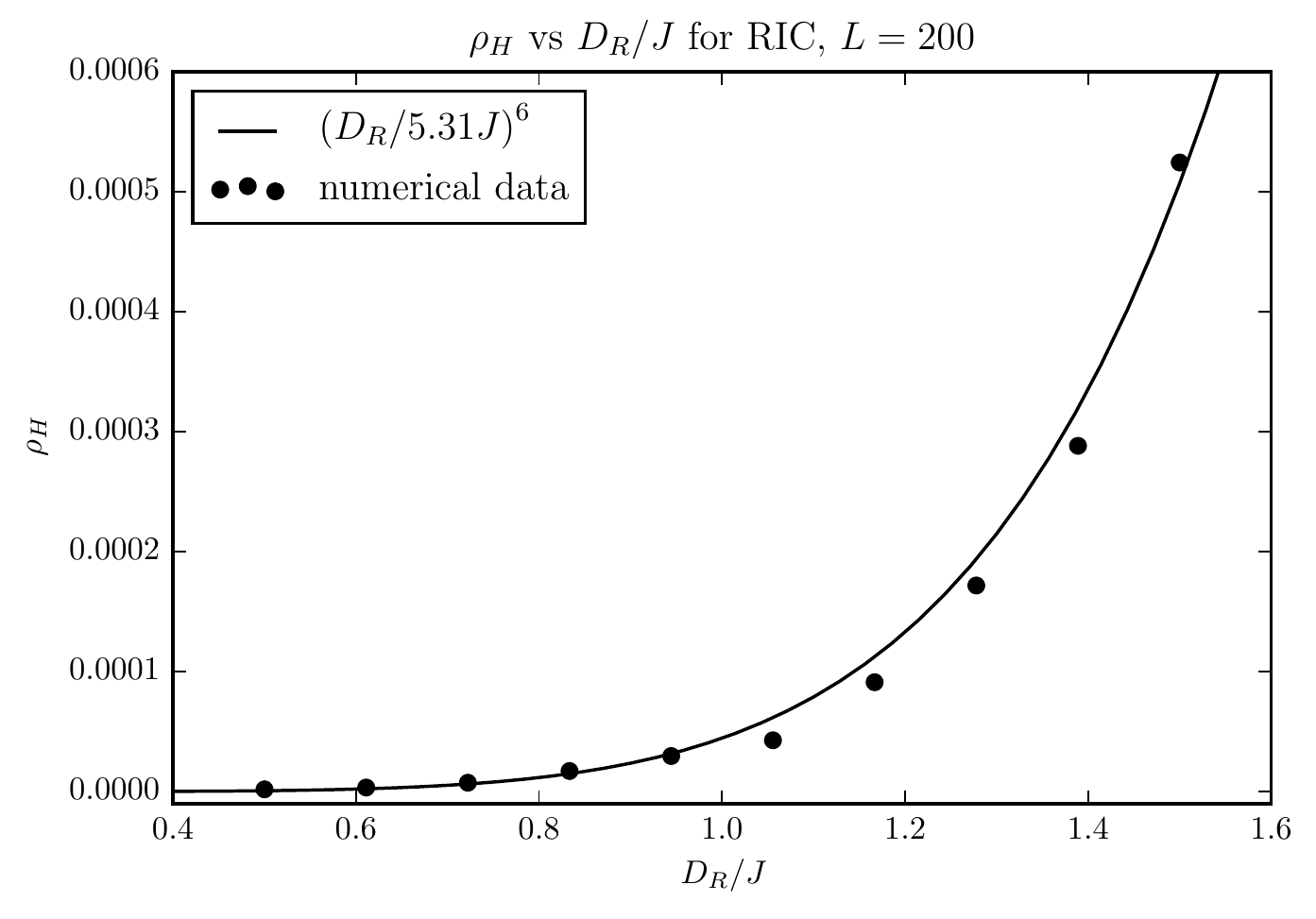}
\caption{\label{fig:HedgeDensity}Hedgehog density vs $D_R$.}
\end{figure}

We can find hedgehogs in our computed states using a simple method: we look for points between lattice sites where spins on opposite sides of the point are aligned in opposite directions. 
This method consistently finds hedgehogs; all other configurations that satisfy this condition are forbidden by theory, so there is no risk for false positives. 
Except for very high strengths of the random anisotropy, collinear initial conditions generally do not produce any singularities. 
Random initial conditions, however, do produce singularities.  
\figref{HedgeDensity} shows the density of hedgehogs, $\rho_H$, i.e.~the ratio between the number of points where hedgehogs have been found and the total number of sites, versus the strength of the random anisotropy, $D_R$.

We have found that $\rho_H \approx \left(0.19 D_R/J\right)^6$. 
Combining this result with \eqnref{Rf}, we obtain 
\begin{equation}
 \rho_H\approx \frac{2.1}{\frac{4}{3}\pi (R_f/a)^3},
\end{equation}
i.e.~there are approximately two Hedgehogs per Imry-Ma domain. 
This finding is in accordance with the topological argument presented a previous work\cite{topology-PRL}: the Imry-Ma state with zero total magnetization requires singularities at $n < d + 1$, where $n$ is the number of spin components and $d$ is dimensionality of space. 

\subsection{Correlated disorder}\label{sintered}
\begin{figure}[h!]
\centering
\includegraphics[width=8.5cm]{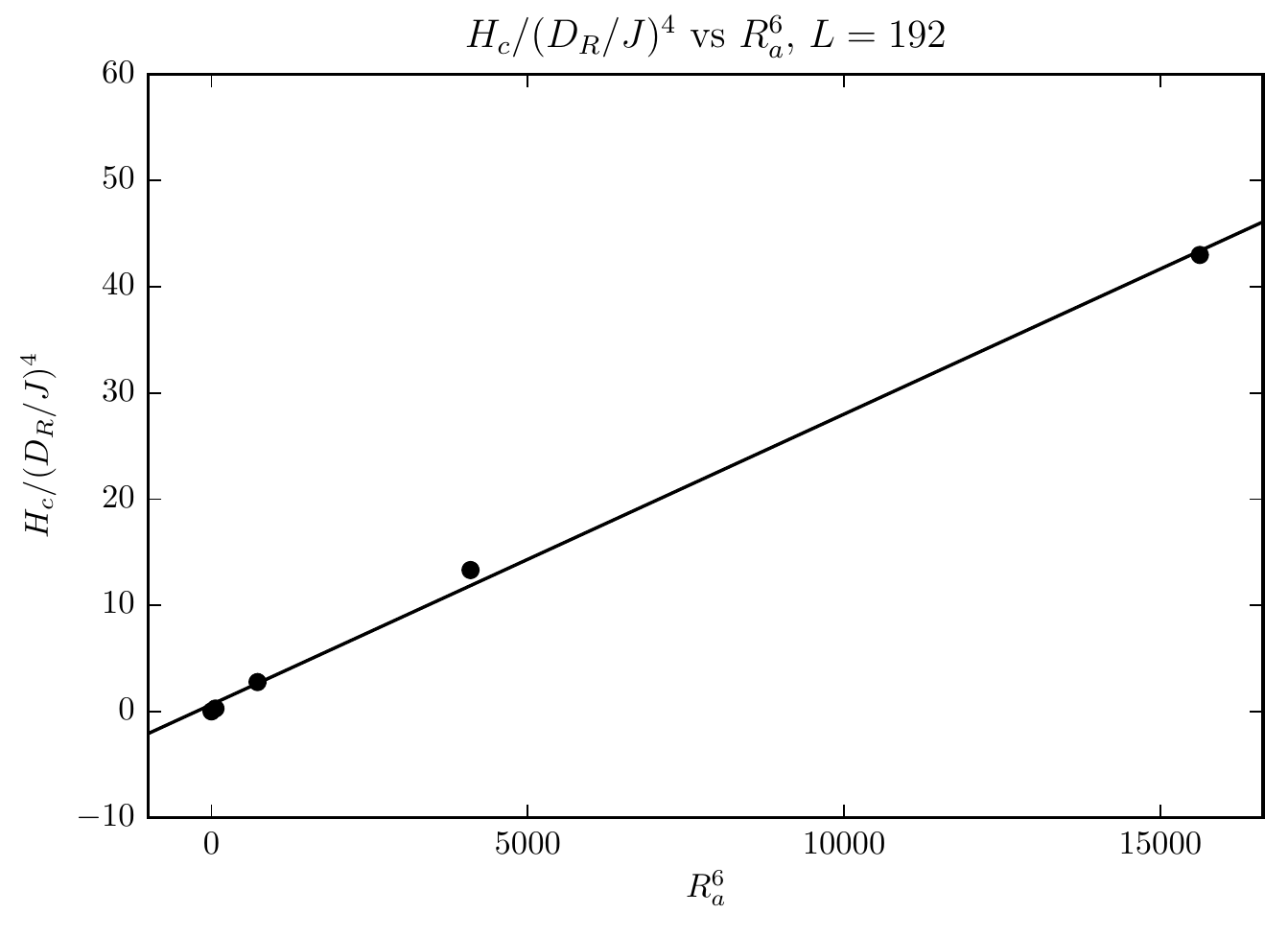}
\caption{\label{fig:correlated}The dependence of the coercive field on the size of the grain, $R_a$. The points correspond to $R_a/a = 1,2,3,4,5$.}
\end{figure}

So far we have studied the site disorder, i.e.~the direction of the anisotropy was chosen randomly at each lattice site. 
Meanwhile, amorphous and sintered magnets would have anisotropy axes correlated on some scale $R_a > a$. This simply replaces $a$ with $R_a$ in \eqnref{Rf}, making 
$R_f/a \propto (a/R_a)^3(J/D_R)^2$ and 
\begin{equation}
H_C \propto D_R\left(\frac{R_a}{R_f}\right)^{3/2} \propto D_R^4 R_a^6, 
\end{equation}
which is valid for $a < R_a \ll R_f \ll L$. 
Under these conditions the sixth power dependence of the coercive field on the grain size is confirmed by numerical results.  
These numerical results are obtained by using cubic correlated chunks, where all sites within a cubic region with volume $R_a^3$ have aligned anisotropy axes.  
This corresponds to the physical conditions in sintered magnets. 
\figref{correlated} shows the dependence of $H_C$ on $R_a$ for $R_a = a, 2a, 3a, 4a, 5a$. 
Note that $R_f$ is proportional to  the inverse third power of $R_a$, which invalidates the condition $R_a \ll R_f$ very fast with increasing $R_a$,  limiting numerical studies of a finite-size system to just a few grain sizes.

\section{Discussion}\label{discussion}
Using $3d$ lattices containing over 10 million spins, we have numerically studied the magnetic properties of the random-anisotropy Heisenberg model in the limit when the anisotropy is sufficiently weak compared to the exchange to provide a ferromagnetic correlation length that is greater than the scale on which anisotropy axes are correlated. 
This limit will be satisfied by many amorphous magnets, as well as by sintered magnets in which the size of the grain, $R_a$, is sufficiently small. Taking $R_f/a = 15\pi (a/R_a)^3(J/D_R)^2$ in accordance with our analytical and numerical results, one obtains that the condition $R_a \ll R_f$ requires $R_a/a \ll (J/D_R)^{1/2}$. 
Had $D_R$ been the magneticrystalline anisotropy, $(J/D_R)^{1/2}a$ would have represented the scale of the domain wall width. 
Consequently, if the magnet was sintered from ferromagetic nanocrystals, the condition $R_a \ll R_f$ would correspond to the condition that the size of the nanocrystal was small compared to the domain wall width in the magnetic material. This is practically feasible and, in fact, reflects the direction in which the magnetic industry is going. 

Under the above condition, we found, in accordance with theoretical expectation, that the coercive field and the area of the hysteresis loop roughly scale as $D_R^4$ and $R_a^6$. 
This strong dependence on parameters shows that decreasing $D_R$ and $R_a$ by even a small factor could drastically reduce the coercive field, paving the way to extremely soft magnetic materials. 
One obstacle could be the coherent anisotropy which is inevitably present in any sample due to its non-spherical shape and/or the anisotropy of the process of sample preparation. Let such anisotropy have strength $D_C$. 
Its effect on the magnetic state will be small if it is weak compared to the effective anisotropy stemming from $D_R$. 
The latter, as our theoretical argument suggests and numerical work confirms, scales as $D_{\rm eff} \sim D_R^4/J^3$. 
Consequently, the condition $D_C \ll D_{\rm eff}$ translates into $D_C/J \ll (D_R/J)^4$. 
Thus, in the case of a weak random anisotropy, a much weaker coherent anisotropy would destroy the softness of the magnet and will convert it into a more conventional ferromagnet with domain walls of the width $\sim (J/D_C)^{1/2}$ pinned by disorder.

An interesting question is the physical origin of metastability. 
We have seen that, similar to the random field model,\cite{topology-PRL} the metastability comes in large part from hedgehogs, whose concentration strongly depends on $D_R$ and corresponds to about two hedgehogs per volume of size $R_f$. 
However, there is a difference from the random-field model. 
In a random field system the metastability and hysteresis disappear for $n > d +1$ when topological defects are absent. 
In contrast, magnetic anisotropy in the random anisotropy model introduces bistability,
  creating topological defects -- domain walls -- regardless of the relation between $n$ and $d$. 
In principle, one can think of domain walls of width $\sim (J/D_{\rm eff})^{1/2}a$. 
However, substitution of $D_{\rm eff} \sim D_R^4/J^3$ into this expression gives a width, $\sim (J/D_R)^2a$, which scales as $R_f$, making the concept of a domain wall separating domains useless. 
Nevertheless, the topology of the random-anisotropy model remains different from the topology of the random-field model. 
We observed this by numerically studying the $3d$ random-anisotropy model with a five-component spin. 
Although the agreement with analytical results becomes more precise when the hedgehogs are absent,  hysteresis persists, unlike the behavior found in the random field model. 

\section{Acknowledgments}

This work was supported by the grant No. DE-FG02-93ER45487 funded by the U.S. Department of Energy, Office of Science.

\bibliography{n=3d=3paper}

\end{document}